\documentclass[conference]{IEEEtran}
 \IEEEoverridecommandlockouts
\usepackage{cite}
\usepackage{amsmath,amssymb,amsfonts}
\usepackage{graphicx}
\usepackage{textcomp}
\usepackage{xcolor}
\usepackage{comment}
\usepackage{optidef}
\usepackage[T1]{fontenc}
\usepackage{mathtools} 
\usepackage{cuted}
\usepackage{flushend}
\usepackage{bbm}
\usepackage{dsfont}
\usepackage{mathtools}
\usepackage{multirow}
\usepackage{float}

\begin{document}
\title{Predictive Resource Allocation for URLLC using Empirical Mode Decomposition
\thanks{This work was supported by the Academy of Finland 6Genesis Flagship (grant no. 346208).}}
 \author{\IEEEauthorblockN{Chandu~Jayawardhana, Thushan~Sivalingam, Nurul~Huda~Mahmood, Nandana~Rajatheva, and Matti~Latva-Aho}
 \IEEEauthorblockA{6G Flagship, Centre for Wireless Communications, University of Oulu, Oulu, Finland}
\{chandu.wijenayakejayawardhana, thushan.sivalingam, nurulhuda.mahmood, nandana.rajatheva, matti.latva-aho\}@oulu.fi
 }
 
\maketitle
\begin{abstract}
\begin{comment}
For future improvements and enhancements of  ultra reliable low latency communication (URLLC) related services, it is essential to simultaneously meet low latency and high reliability requirements. Reducing retransmissions of URLLC packets is one way to achieve high reliability while minimising latency~\cite{693631}. To reduce the number of retransmissions, resources must be efficiently allocated for desired link to achieve the reliability target at single shot transmission. In order to allocate resources efficiently having knowledge about the interference is essential. This paper proposes empirical mode decomposition based hybrid prediction method to predict the interference and resource allocation for downlink based on the predicted values. The proposed method is found to meet near optimal allocation of resources and evaluated based on baseline criteria. 
\end{comment}
Effective resource allocation is a crucial requirement to achieve the stringent performance targets of ultra-reliable low-latency communication (URLLC) services. Predicting future interference and utilizing it to design efficient interference management algorithms is one way to allocate resources for URLLC services effectively. This paper proposes an empirical mode decomposition (EMD) based hybrid prediction method to predict the interference and allocate resources for downlink based on the prediction results. EMD is used to decompose the past interference values faced by the user equipment. Long short-term memory and auto-regressive integrated moving average methods are used to predict the decomposed components. The final predicted interference value is reconstructed using individual predicted values of decomposed components. It is found that such a decomposition-based prediction method reduces the root mean squared error of the prediction by $20 - 25\%$. The proposed resource allocation algorithm utilizing the EMD-based interference prediction was found to meet near-optimal allocation of resources and correspondingly results in $2-3$ orders of magnitude lower outage compared to state-of-the-art baseline prediction algorithm-based resource allocation. 

\end{abstract}
\vspace{3mm}

\section{Introduction}
The ultra-reliable low-latency communication (URLLC) service class has been introduced in fifth-generation (5G) new radio to support mission-critical applications where uninterrupted and robust data exchange is of the utmost importance. It has use cases in factory automation, process industries, smart grids, intelligent transport systems, and professional audio applications~\cite{03}. The latency and reliability targeted by URLLC services are one millisecond and $1 - 10^{-5}$, respectively~\cite{04}.

URLLC is confronted with three significant lower layer challenges: uncertain traffic arrival, random channel impairments including fading, and random interference~\cite{9245553}. Conventional resource allocation methods address these challenges by adopting the transmission strategy based on the observed outcome, for instance, hybrid automatic repeat request (HARQ) re-transmissions in the case of a transmission failure~\cite{8490699}. As the latency and reliability requirements for URLLC services are much more stringent, such conventional schemes are not able to support it efficiently~\cite{8472907}.

Existing 5G URLLC enabling technologies separately focus on achieving low latency and high reliability. Although these independently created reliability and latency-enabling technologies meet the 5G standards, these existing technologies may not be capable enough to meet the requirements of networks beyond 5G. For example, 5G achieves reliability requirements by over-provisioning resources, which otherwise could have been allocated for some other service, such as enhanced mobile broadband (eMBB) or massive machine-type communication (mMTC). Therefore, an innovative, intelligent solution must simultaneously address latency and reliability while utilizing scarce resources efficiently.

Developing an intelligence mechanism to predict future interference values and allocate resources proactively is seen as a potential URLLC enabler that can facilitate efficient resource allocation~\cite{9838521}. Link adaptation is a traditional method that predicts future interference values based on past interference samples~\cite{7962790}. However, one drawback of conventional link adaptation schemes is that they only consider the mean value of all interference values from the past, ignoring the sudden interference fluctuations. This behavior may result in inaccurate predictions because it lacks knowledge of the entire interference distribution. Hence, an interference prediction scheme that utilizes the whole past interference distribution for the forecast will lead to better predictions. %. If we could decompose this random signal into a set of correlated signals, it would create a different perspective on investigating the interference in detail. Consequently, the interference prediction for each decomposed signal would be less complex and more elaborate.

A recursive predictor is presented in~\cite{9354031}, where the autocorrelation of interference is converted into an autoregressive moving average representation, which is then utilized to estimate future interference. Predictive interference-aware resource allocation for graphical processor unit systems for efficiently training deep learning models are presented in~\cite{9428512}. In addition, interference prediction for vehicular communication is proposed in~\cite{9759361}, where packet collision is avoided by the knowledge of the predicted interference vehicles. Since interference is a highly random signal with contributions from multiple interferers, accurately predicting interference is a complex procedure. Hence, more accurate prediction mechanisms need to be explored. Towards this end, we propose a novel empirical mode decomposition (EMD) based approach.
\begin{comment}
Add here one paragraph with a summary of a few current literature and how they approached this problem of interference prediction. (some examples: https://ieeexplore.ieee.org/document/9354031 
https://ieeexplore.ieee.org/document/9759361
https://ieeexplore.ieee.org/document/9428512)
Then say that: 
Since interference is a highly random signal with contributions from multiple interferers, predicting it accurately is a complex procedure. Hence, more accurate prediction mechanisms need to be explored. Towards this end, we propose a novel empirical mode decomposition (EMD) based approach.
\end{comment}

The study of efficient resource allocation for URLLC services based on predictive interference values is still in the early stages~\cite{8879655}. Therefore, we propose a novel hybrid resource allocation scheme using EMD~\cite{5596829}. In the proposed method, we decomposed the original signal using EMD and predicted the decomposed signal components using long-short term memory (LSTM)~\cite{12} and autoregressive integrated moving average (ARIMA)~\cite{08}. Then, individually predicted components are added to form a total interference prediction. Finally, resources are efficiently allocated in the downlink based on the predicted interference values. The proposed scheme achieves near-optimal performance, which is difficult to achieve using conventional interference prediction and resource allocation methods.

The rest of this paper is organized as follows. Section~\ref{sec:systemModel} represents the system model and problem statement. In Section~\ref{sec:Int_pred_res_alloc}, we propose the EMD-based prediction and resource allocation scheme. Simulation results and performance analysis are given in Section~\ref{sec:sim_results}. Finally, conclusions are presented in Section~\ref{sec:conclu}.

\section{System Model}
\label{sec:systemModel}

\begin{figure}[t]
    \center
    \includegraphics[width=0.6\linewidth]{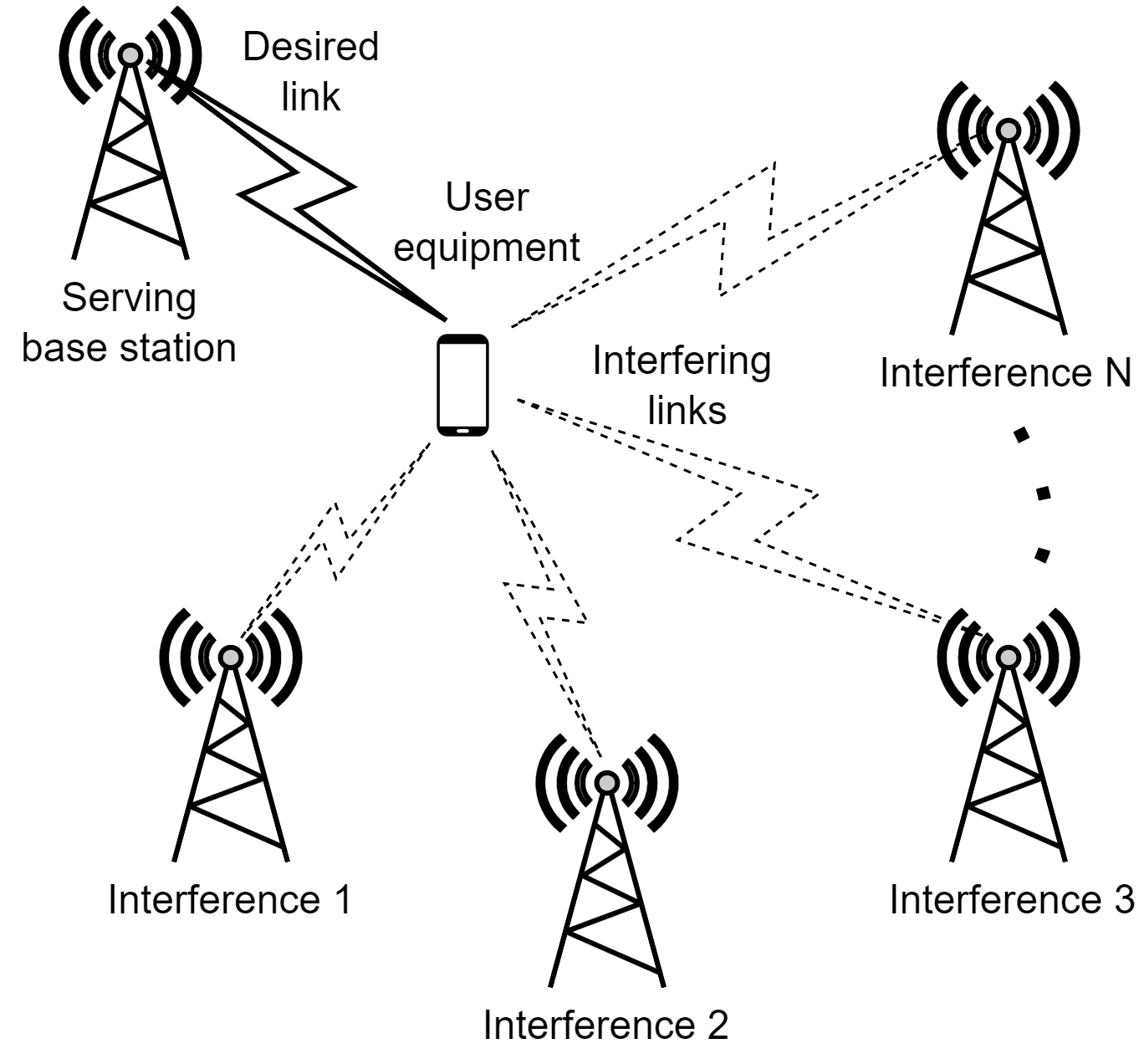}
    \caption{System model: desired link operates with $N$ interferers.}
    \label{fig:system_model}
\end{figure}

Consider the downlink of a wireless network. We are interested in the performance of a URLLC user operating in the presence of $N$ interferers, as shown in Fig.~\ref{fig:system_model}. It is assumed that the desired channel has an average signal-to-noise ratio (SNR) of $\bar{\gamma}$, whereas the mean SNR of the interference links (termed as interference-to-noise ratios (INR)) are uniformly distributed within a given range $[{\gamma}_{min}, {\gamma}_{max}]$. %The above mentioned INR can also be referred to as the SNR of an interference signal.
\begin{comment}
In this study, the downlink of a wireless network is examined. The objective is to determine the optimal resource allocation for a URLLC service operating in the presence of $N$ interferers, as shown in Fig. \ref{fig:system_model}. It is assumed that the desired channel has a signal-to-noise ratio (SNR) of $\bar{\gamma}$ and the interference-to-noise ratios (INR) for each interfering link are uniformly distributed within a given range $[{\gamma}_{min}, {\gamma}_{max}]$. The above mentioned INR can also be referred to as the SNR of an interference signal.
\end{comment}
The desired link is assumed to be established with the base station, which has the highest SNR. Therefore, it is assumed that the interfering link with the highest SNR value ($\gamma_{max}$) has a lower SNR value compared to the desired link ($\bar{\gamma} > {\gamma}_{max}$). Also, we assume there is no cooperation between transmitters. Further, we consider a single antenna (single input single output (SISO)) Rayleigh block fading model. The serving base station transmits $D$ bits via the desired downlink with the given target error probability $\varepsilon$. The objective is to allocate channel uses $R$ as efficiently as possible to guarantee both low latency and high reliability by predicting future interference values. 

URLLC transmissions typically occur over a short transmission time interval (TTI) that can be as low as $0.1$ ms~\cite{8329620}. Since the coherence time of a typical wireless transmission is much longer than this duration, the channel coherence time can span over multiple TTIs. It is assumed that the transmitter acquires sufficient knowledge about the desired channel state information (CSI) within that coherence time. Also, we consider the user equipment (UE) will update the serving base station with the total interference values it has seen in the past. This will give the serving base station more information about the interference at the receiver, enabling it to perform the proposed interference prediction scheme in the downlink. % lead to a more efficient allocation of resources in the downlink.

\section{Interference Prediction and Resource Allocation}
\label{sec:Int_pred_res_alloc}
\begin{figure*}[tb]
    \center
    \includegraphics[width=0.6\linewidth]{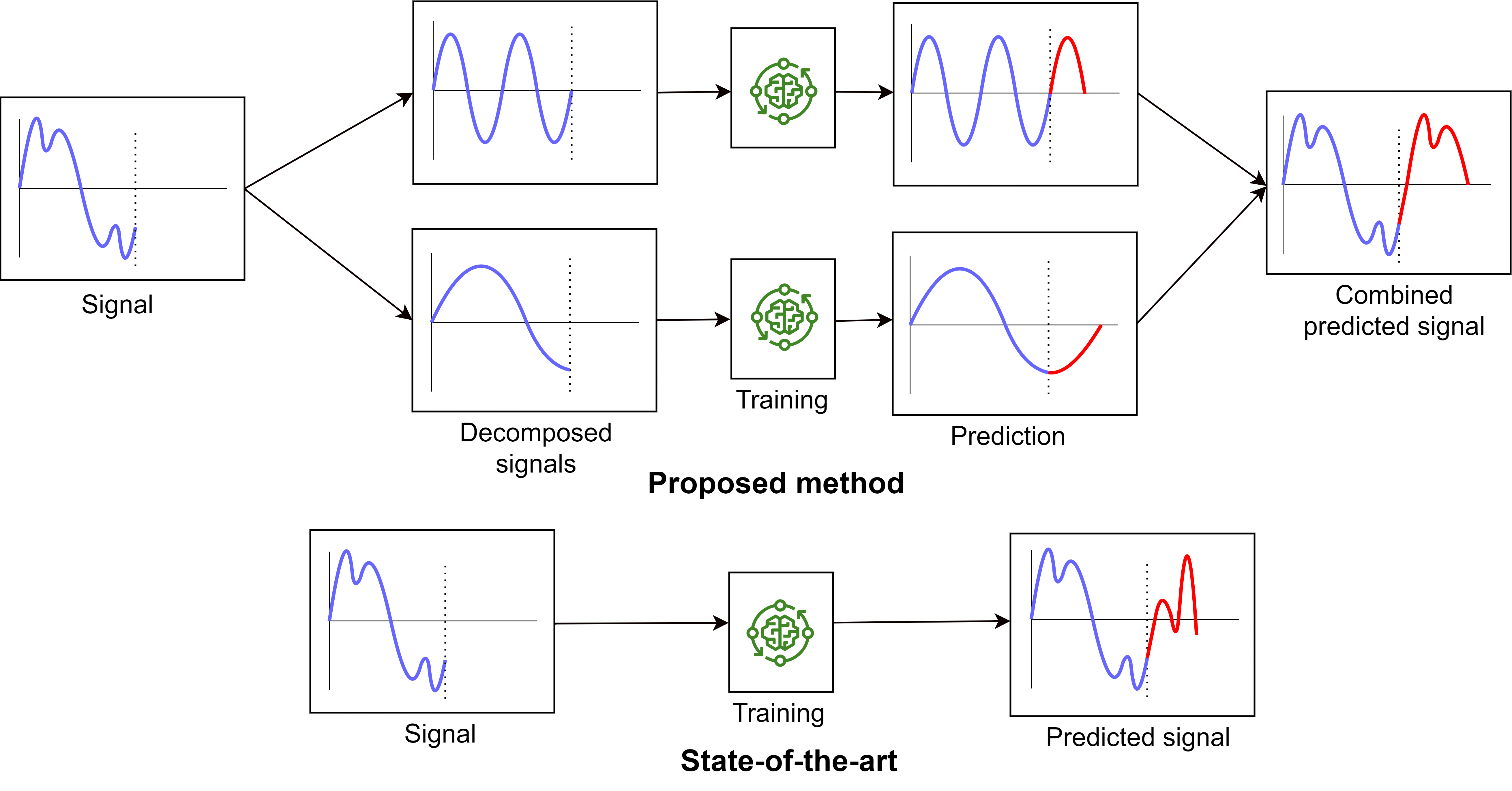}
    \caption{Proposed resource allocation scheme versus state of the art resource allocation scheme.}
    \label{fig:proposed_model}
\end{figure*}

In this work, we propose decomposing the complex sum interference signal using EMD and predicting the individual components separately to improve the prediction accuracy. A schematic of the proposed method and state-of-the-art is shown in Fig.~\ref{fig:proposed_model}. 

\begin{comment} A schematic of the proposed prediction method is shown in Fig. \ref{fig:predict_mode}. 
\begin{figure}[t]
    \center
    \includegraphics[width=\linewidth]{each_block.png}
    \caption{Proposed method for interference prediction.}
    \label{fig:predict_mode}
\end{figure}
EMD based hybrid prediction method is used with two prediction methods LSTM and ARIMA. Resources are efficiently allocated using finite block length theory relying on the predicted interference values. The total interference signal power that the mobile station experiences is derived by adding all the individual interference signal powers as
\begin{equation}
\label{eqn:total_interference}
{f}(t) =\sum_{i=1}^{N} {Interference}{(i)},
\end{equation}
where ${Interference}{(i)}$ is the interference signal power from the $i$\textsuperscript{th} interferer and ${f}(t)$ is the total interference power observed by the UE.
\end{comment}

\subsection{Empirical mode decomposition}

 EMD decomposes the past total interference signal power into intrinsic mode functions (IMF) using the heuristic decomposition method~\cite{5596829}. As a result of this decomposition method, we can obtain frequency-ordered IMF components. Initial IMFs contain high-frequency oscillations. Then, the oscillation frequency reduces as the IMF number increases, resulting in more linear IMFs at the end. The total interference signal power can be reconstructed using decomposed components as
 \begin{equation}
\label{eqn:total}
{f}(t)=\sum_{i=1}^{L-1} {IMF}_{i}(t) + {res}(t),
\end{equation}
where $L$ is the number of decomposed components, and ${res}(t)$ is the residual. Here, $L$ is not predetermined, and it only depends on the nature of the signal. Prediction of a high oscillatory signal is more challenging than a less oscillatory signal when developing and training a model. As a result, instead of relying on a single model for overall forecasting interference, we have devised separate models for each IMF and a residual component. Here, the input signal to each model is less oscillatory as the IMF number increases. Therefore, the model's prediction accuracy improves, and the model training task becomes more manageable. Similar to \eqref{eqn:total}, the predicted value of total interference can also be reconstructed by taking the individual predicted values of each IMFs and residual.

The total interference signal is split into two sets to obtain training and validation samples. This split can be done using a predetermined training sample percentage value (e.g., $80\%$). The number of samples in the total interference signal is denoted as $T$. The number of samples in training data and validating data are taken as $P$ and $M$, respectively, where $T = P+M$.

\subsection{Prediction model selection and training}
Two time series forecasting methods are used to predict the values of each decomposed component, namely LSTM~\cite{08} and ARIMA~\cite{12}. The ARIMA model is selected due to its ability to work with non-stationary data~\cite{12}. The process of ARIMA consists of four steps: identification, estimation, diagnostics, and forecasting. The LSTM recurrent neural network (RNN) method is selected due to its capability of learning long-term dependencies and carrying forward learned information for a long period by supporting future decisions~\cite{08} by addressing the drawbacks of traditional RNN such as vanishing or exploding gradients. 

 Each IMF and the residual are fed into these two forecasting algorithms. Here models have been trained individually corresponding to each IMF and residual with $P$ number of training samples to optimize the prediction accuracy of each component signal. The total number of models trained using the LSTM method equals adding the number of IMFs and residuals ($L$), which is the same as ARIMA. These two forecasting methods predict the next sample value based on the trained neural networks and time series forecasting techniques. 
 
Since there are $M$ validating data points, the one-step prediction method is repeated $M$ times to obtain the predicted value set. The advantage of iteratively predicting only one future value at a time over predicting all $M$ future values at once is that we can optimize the model at each step by comparing the most recent predicted value to the actual value. We can improve prediction accuracy by always incorporating the most recent data points into model training. The forecasted IMF and residual values of each model are then added to obtain the final forecasted interference signal using two methods as
 \begin{equation}
\label{eqn:LSTM_forecast}
{LSTM}_{recon}=\sum_{t=1}^{M}({\sum_{i=1}^{L-1}{IMF}_{LSTM(i)}{(t)}+{res}_{LSTM}{(t)}}),
\end{equation}
\begin{equation}
\label{eqn:ARIMA_forecast}
{ARIMA}_{recon}=\sum_{t=1}^{M}({\sum_{i=1}^{L-1}{IMF}_{ARIMA(i)}{(t)}+{res}_{ARIMA}{(t)}}),
\end{equation}
where ${LSTM}_{recon}$ is the reconstructed signal using the prediction of each IMF and residual using the LSTM method, ${ARIMA}_{recon}$ is the reconstructed signal using the prediction of each IMF and residual using ARIMA method. ${IMF}_{LSTM(i)}{(t)}$ is the predicted value using LSTM model that trained to predict IMF($i$) at time step $t$, ${IMF}_{ARIMA(i)}{(t)}$ is the predicted value using ARIMA model that trained to predict IMF($i$) at time step $t$. Here, ${res}_{LSTM}{(t)}$, ${res}_{ARIMA}{(t)}$ are the residual value of the LSTM model, ARIMA model at time step $t$, respectively. 

In addition to the decomposed signals, training samples of the total interference are fed into the LSTM and ARIMA models for baseline comparison. The corresponding predictions are obtained as
\begin{equation}
\label{eqn:LSTM_signal}
{LSTM}_{Signal} =\sum_{t=1}^{M}{Signal}_{LSTM}{(t)},
\end{equation}
\begin{equation}
\label{eqn:ARIMA_signal}
{ARIMA}_{Signal} =\sum_{t=1}^{M}{Signal}_{ARIMA}{(t)},
\end{equation}
where ${LSTM}_{Signal}$ is the predicted signal using LSTM method without using EMD, ${ARIMA}_{Signal}$ is the predicted signal using ARIMA method without using EMD, ${Signal}_{LSTM}{(t)}$ is the predicted value at time step $t$ using LSTM method, and ${Signal}_{ARIMA}{(t)}$ is the predicted value at time step $t$ using ARIMA method.

\subsection{Resource allocation}
\begin{figure}[t]
    \center
    \includegraphics[width=0.9\linewidth]{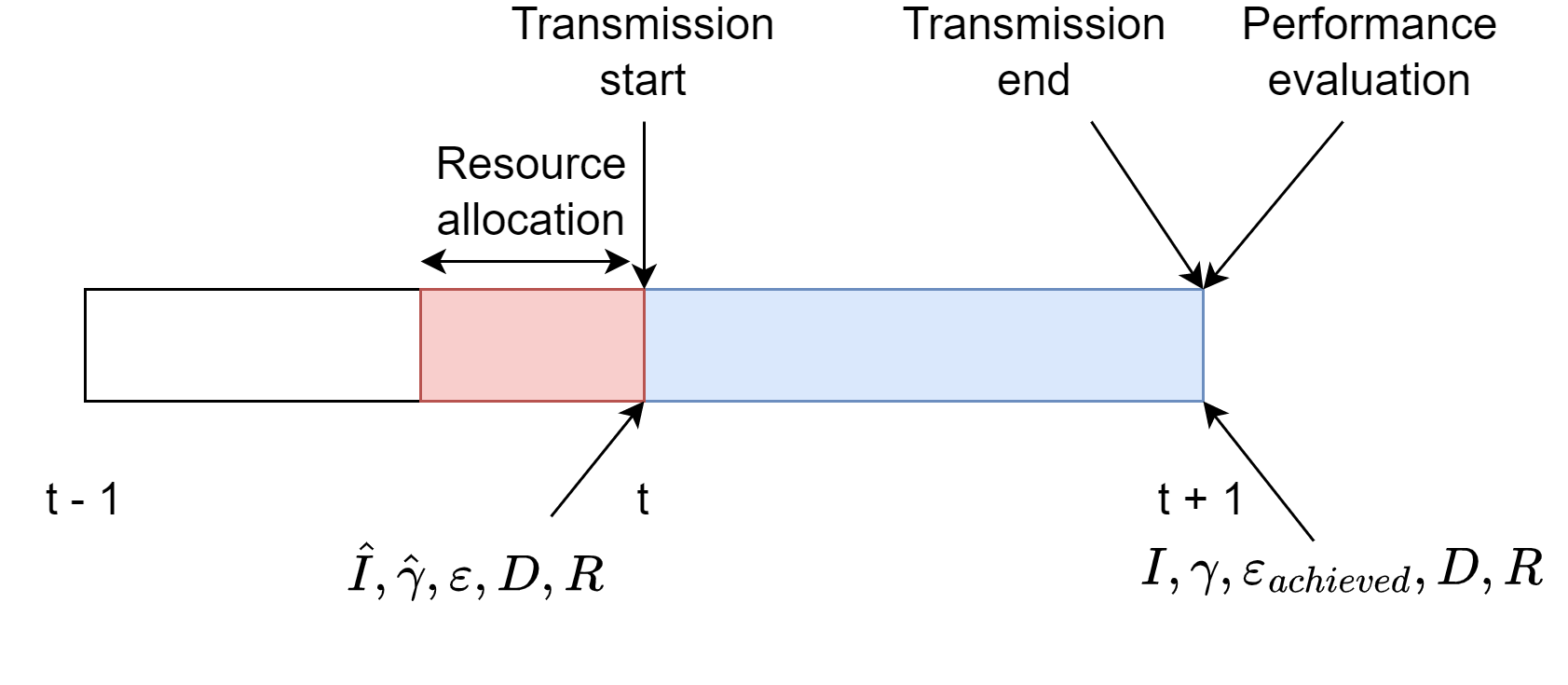}
    \caption{Resource allocation and performance evaluation timeline.}
    \label{fig:timeline}
\end{figure}

The predicted signal to interference and noise ratio (SINR) $\hat{\gamma}$ will be calculated assuming that the power of the desired signal $S$ is known using CSI estimates, and is given by
\begin{equation}
\label{eqn:SINR_hat}
\hat{\gamma} = \frac{S}{\hat{I} + N_{0}},
\end{equation}
where $\hat{I}$ is the predicted interference power and $N_{0}$ is the normalized noise power. According to finite block length theory~\cite{09}, the number of information bits $D$ that can be transmitted with the given decoding error probability $\varepsilon$ using $R$ channel uses in additive white gaussian noise (AWGN) channel can be calculated by
\begin{equation}
\label{eqn:Number_of_bits}
D = RC(\hat{\gamma}) - Q^{-1}(\varepsilon)\sqrt{RV(\hat{\gamma} )} + O(\log_{2}R),
\end{equation}
where $C(\hat{\gamma} ) = \log_{2}(1 + \hat{\gamma})$ is the shannon capacity of AWGN under a finite block length regime, $V(\hat{\gamma} ) = \frac{1}{ln(2)^{2}}\left( 1 - \frac{1}{(1+\hat{\gamma} )^{2}} \right)$ is the channel dispersion measured in squared information units per channel, and $Q^{-1}$ is the inverse of the Q-function~\cite{9245553}. Using the preceding information, the number of resources (channel utilisation) $R$ required to transmit $D$ bits with a target error probability $\varepsilon$ is estimated by~\cite{8490699}

\begin{equation}
\label{eqn:channel_usage}
R \approx {\frac{D}{C(\hat{\gamma} )} + \frac{Q^{-1}(\varepsilon)^{2}V(\hat{\gamma} )}{2C(\hat{\gamma} )^{2}}  \left[1 + \sqrt{1 + \frac{4DC(\hat{\gamma} )}{Q^{-1}(\varepsilon)^{2}V(\hat{\gamma} )}}  \right]}.
\end{equation}
This resource allocation happens within one time step prior to the actual transmission ($t-1$), as shown in Fig.~\ref{fig:timeline}.
After the transmission of $D$ bits using the allocated $R$ number of resources at time $t$, we can obtain the exact total interference $I$ that $D$ bits have undergone. Therefore, we can obtain the actual SINR $\gamma$ as shown in \eqref{eqn:SINR}. It is important to note that, now $\gamma$ will be calculated using actual total interference $I$ value as
\begin{equation}
\label{eqn:SINR}
\gamma = \frac{S}{I + N_{0}}.
\end{equation}
Since we initially allocate the resources $R$ based on target error probability $\varepsilon$, now we can  obtain the achieved error probability by rearranging the terms in \eqref{eqn:Number_of_bits} and substituting predicted SINR value $\hat{\gamma}$ with the actual SINR value $\gamma$ as shown below

\begin{equation}
\label{eqn:decoding_error_actual}
{\varepsilon}_{achieved} \approx{Q\left(\frac{RC(\gamma )-D}{\sqrt{RV(\gamma )}}\right)}.
\end{equation}
Finally, we will be able to compare the target error probability with the achieved error probability. The above-mentioned procedure can be iterated over different target error rate values to obtain the variation of achieved error rate values against the target error rate values.

\section{Simulations Results}
\label{sec:sim_results}
Unless otherwise specified, the simulation parameters are listed in Table \ref{tab1}. The simulation parameters can be changed based on requirements without compromising the model's performance.
\begin{table}[htbp]
\caption{Simulation parameters}
\begin{center}
\begin{tabular}{ll}
\hline
\textbf{Description}&\textbf{Value} \\
\hline
Number of interferers ($N$) & 5 \\
SNR values of interferers & [5, 3, 0, -2, -5] dB \\
SINR value of desired signal ($\bar{\gamma}$) & 20 dB \\
Number of samples considered & 100 \\
Channel model & Rayleigh block fading SISO\\
Number of Bits ($D$) & 50\\
Target error rates ($\varepsilon$) & $10^{-5},10^{-4},10^{-3},10^{-2},10^{-1}$ \\
Forgetting Factor of IRR ($\alpha$) & 0.01\\
\hline
\end{tabular}
\label{tab1}
\end{center}
\end{table}

\subsection{EMD based prediction accuracy analysis} \label{EMD_prediction}
We used the same hyperparameter settings for all models to validate the EMD-based prediction accuracy in comparison to conventional prediction methods that do not use the EMD. Although the prediction models are trained according to individual training datasets of IMFs, residual, and total signal, the hyperparameters used for all these models remain the same. All the models are trained using the following hyperparameters in Table \ref{tab2}.%For example, all the models are trained using the following hyper parameters given in Table \ref{tab2} and obtained individual predictions as shown in Fig. \ref{fig:training_model_select}.
\begin{table}[htbp]
\caption{Hyperparameters used to train all models}
\begin{center}
\begin{tabular}{lll}
\hline
\textbf{Method}&\textbf{Hyperparameter}&\textbf{Value} \\
\hline
LSTM/ARIMA & Training window & 30 \\
\hline
\multirow{7}{*}{LSTM} & Number of epochs & 100 \\
& Activation function & ReLU \\
& Optimizer function & Adam\\
& Loss function & MSE \\
& Number of neurons in LSTM layer 1 & 100 \\
& Number of neurons in LSTM layer 2 & 100 \\
& Number of neurons in dense layer & 1 \\
\hline
\multirow{3}{*}{ARIMA}& Number of previous samples for AR ($p$) & 30 \\
& Number of previous samples for MA ($q$) & 0 \\
& Order of integration ($d$) & 1 \\
\hline
\end{tabular}
\label{tab2}
\end{center}
\end{table}

\begin{figure}[t]
    \center
    \includegraphics[width=0.85\linewidth]{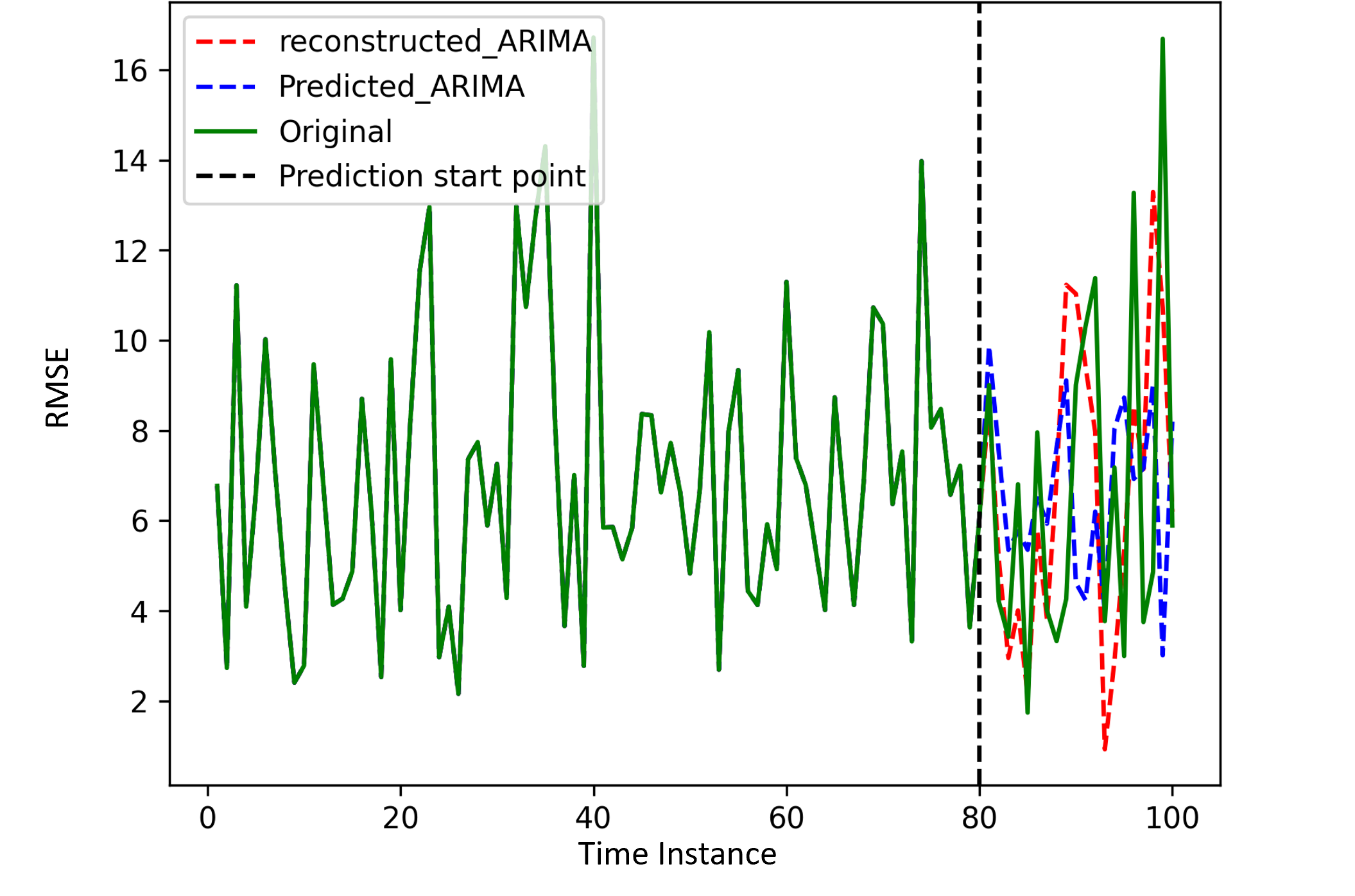}
    \caption{RMSE comparision of ARIMA method.}
    \label{fig:compare_ARIMA}
\end{figure}

\begin{figure}[t]
    \center
    \includegraphics[width=0.85\linewidth]{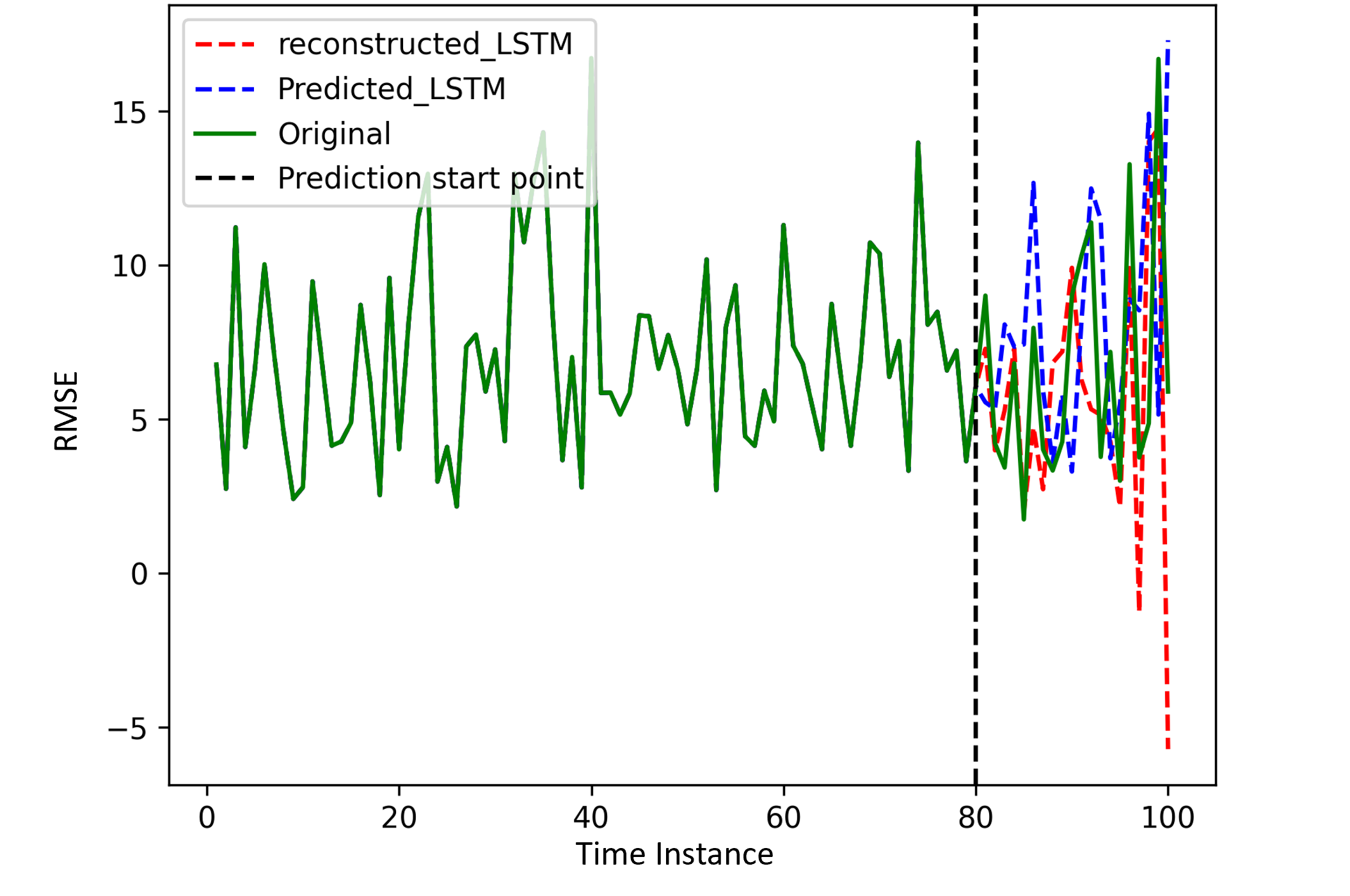}
    \caption{RMSE comparision of LSTM method.}
    \label{fig:compare_LSTM}
\end{figure}

Individually predicted elements will be added to generate the EMD-based prediction output. The performances of EMD-based prediction methods are compared with conventional prediction methods (without EMD). We use the root mean squared error (RMSE) as an evaluation criterion to compare the performance. In each method, an RMSE value is obtained by comparing the reconstructed signal with the validation dataset of the total interference signal. Then, an RMSE value is again obtained by comparing the predicted value of the interference signal with the validation dataset of the total interference signal. The obtained RMSE values are presented in Table \ref{tab3}, and predicted signals are shown in Fig.~\ref{fig:compare_ARIMA} and Fig.~\ref{fig:compare_LSTM}.

\begin{table}[htbp]
\caption{RMSE comparison of the predicted signals with the original signal}
\begin{center}
\begin{tabular}{lll}
\hline
\textbf{Comparison criteria}&\textbf{ARIMA}&\textbf{LSTM} \\
\hline
Original with predicted (without EMD) & 2.13 & 2.49 \\
Original with reconstructed (with EMD) & 1.63 & 1.91 \\
\hline
\end{tabular}
\label{tab3}
\end{center}
\end{table}
As shown in Table \ref{tab3}, the RMSE values of the reconstructed signals using both LSTM and ARIMA methods are less than the RMSE values of the predicted signals. Therefore, it suggests that EMD-based prediction methods outperform conventional prediction methods.
\subsection{Baseline Schemes for resource allocation}
We evaluate the performance of the proposed scheme in comparison to the baseline. Therefore, the following two baseline schemes have been used~\cite{09}.
% \begin{itemize}
% \item \textbf{Moving Average Based Estimation}
\subsubsection{Moving Average Based Estimation}
This is a conventional weighted average-based estimation procedure. It has been adapted to predict interference signal power as the first baseline scheme. Here, interference measured at time $t$ is passed through the first order, low pass, infinite impulse response (IIR) filter. Interference estimates are obtained as 
\begin{equation}
\label{eqn:IIR_filter}
\hat{I}_{t+1} = \alpha I_{t-1} + (1 - \alpha)\hat{I}_{t}, 
\end{equation}
where, $\alpha$ $(0<\alpha<1)$ is the forgetting factor (FF) of the filter. It determines the weight given to the latest measurement compared to the previous one. Based on a simulation-based heuristic analysis, it has been determined that an $\alpha = 0.01$ is optimal for latency performance~\cite{7962790}.

% \item \textbf{Genie Aided Estimation}
\subsubsection{Genie Aided Estimation}
The genie-aided estimator is considered an optimal estimator. In this interference prediction scenario, it is assumed that the transmitter knows precisely what interference condition the transmitted signal will encounter. In other words $\hat{I} = I$.

For comparison, we evaluate the performance of the proposed scheme with the baseline schemes, where the predicted interference values using baseline schemes are $\hat{I}_{IIR}$ and $\hat{I}_{Genie}$. Then, achieved outages for the same set of target outage values will be obtained using the procedure described above. Finally, conclusions were made by comparing the results of the proposed scheme to the results of the baseline schemes.

\subsection{Overall resource allocation and performance analysis}
We used the same hyperparameters for all models in section \ref{EMD_prediction} only to compare the prediction accuracy of EMD-based predictions with the conventional prediction methods. We can achieve high prediction accuracy by individually tuning hyperparameters for each model because the characteristics of the input signal to each model vary as the IMF number increases. Suppose we use the same hyperparameters for each model. In that case, the models with more linear input signals will encounter overfitting, and the models with more random input signals will experience underfitting. Also, the model training time can be drastically reduced by using relaxed hyperparameters for higher IMFs. For example, using a small training window for higher IMFs will result in less training time and still be able to obtain highly accurate predictions.

Resources are allocated for the downlink channel based on predicted interference $\hat{I}$. Achieved block error rate targets ${\varepsilon}_{achieved}$ were obtained by simulating different target error rate values $\varepsilon$. The performance of the EMD-based resource allocation is evaluated against the two baseline schemes.
\begin{figure}[t]
    \center
    \includegraphics[width=0.85\linewidth]{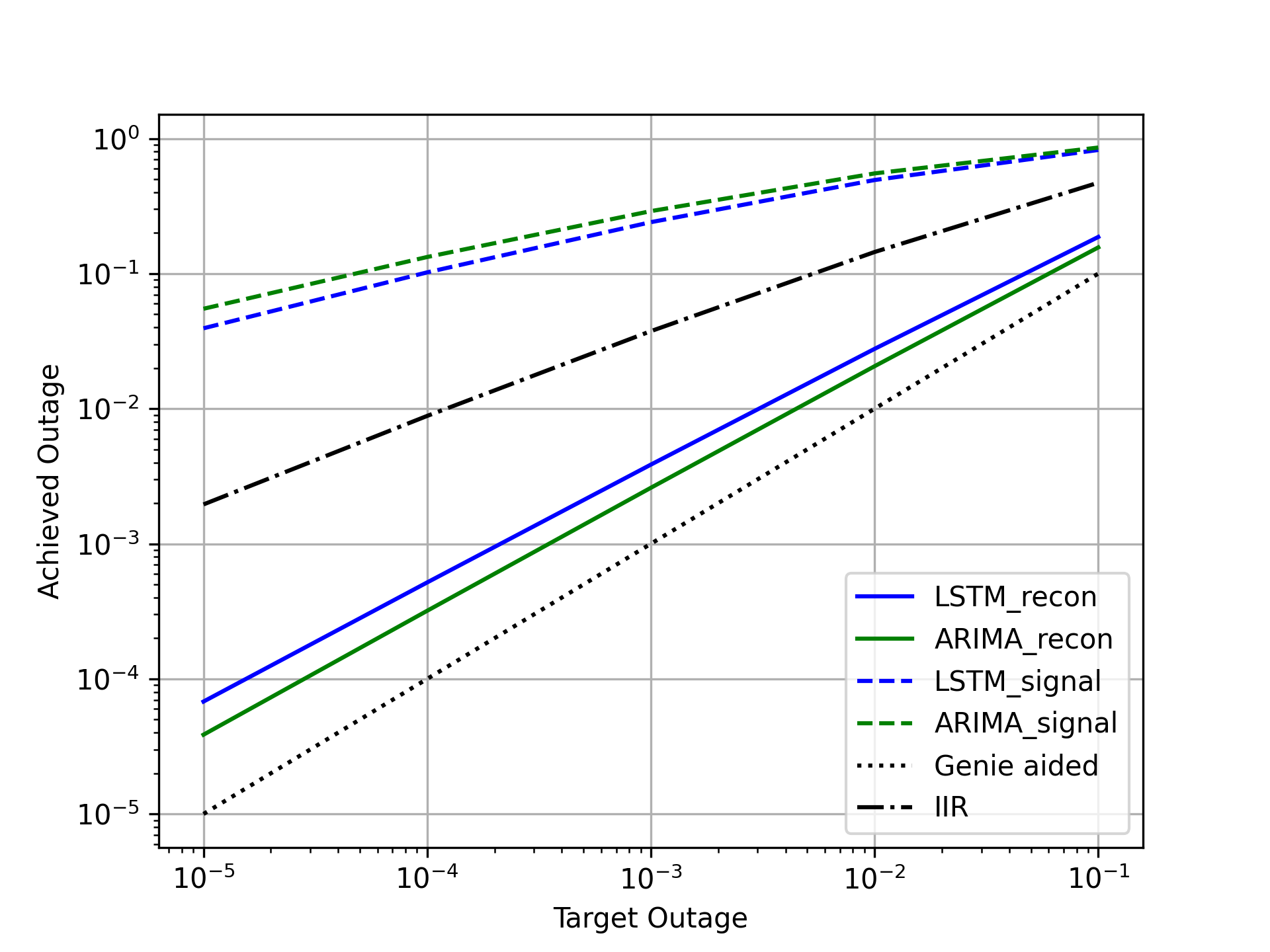}
    \caption{Achieved outage versus target outage for different prediction methods.}
    \label{fig:resource_alloc}
\end{figure}
Since the genie-aided estimator assumed it already knows the achieved SINR, it can allocate the exact number of resources required to achieve the target outage, as shown in Fig.~\ref{fig:resource_alloc}. The curve generated by the Genie-aided estimator is regarded as the optimal allocation of resources. Any curve that deviates less from the genie-aided curve can be considered an efficient allocation of resources. In contrast, efficiency decreases as the curve moves further from the genie-aided curve.

The performance of the IIR filter-based estimation shows low performance and can only achieve the block error rate (BLER) target of around $10\%$, as shown in Fig.~\ref{fig:resource_alloc}. IIR filter-based achieved outage curve shows considerable deviation from the expected genie aided achieved outage curve. The achieved BLER targets of IIR filter-based estimation can be quite resource-efficient only for eMBB services~\cite{7414384}. For URLLC services, stricter achieved outage targets are anticipated.
 
\begin{figure}[t]
    \center
    \includegraphics[width=0.85\linewidth]{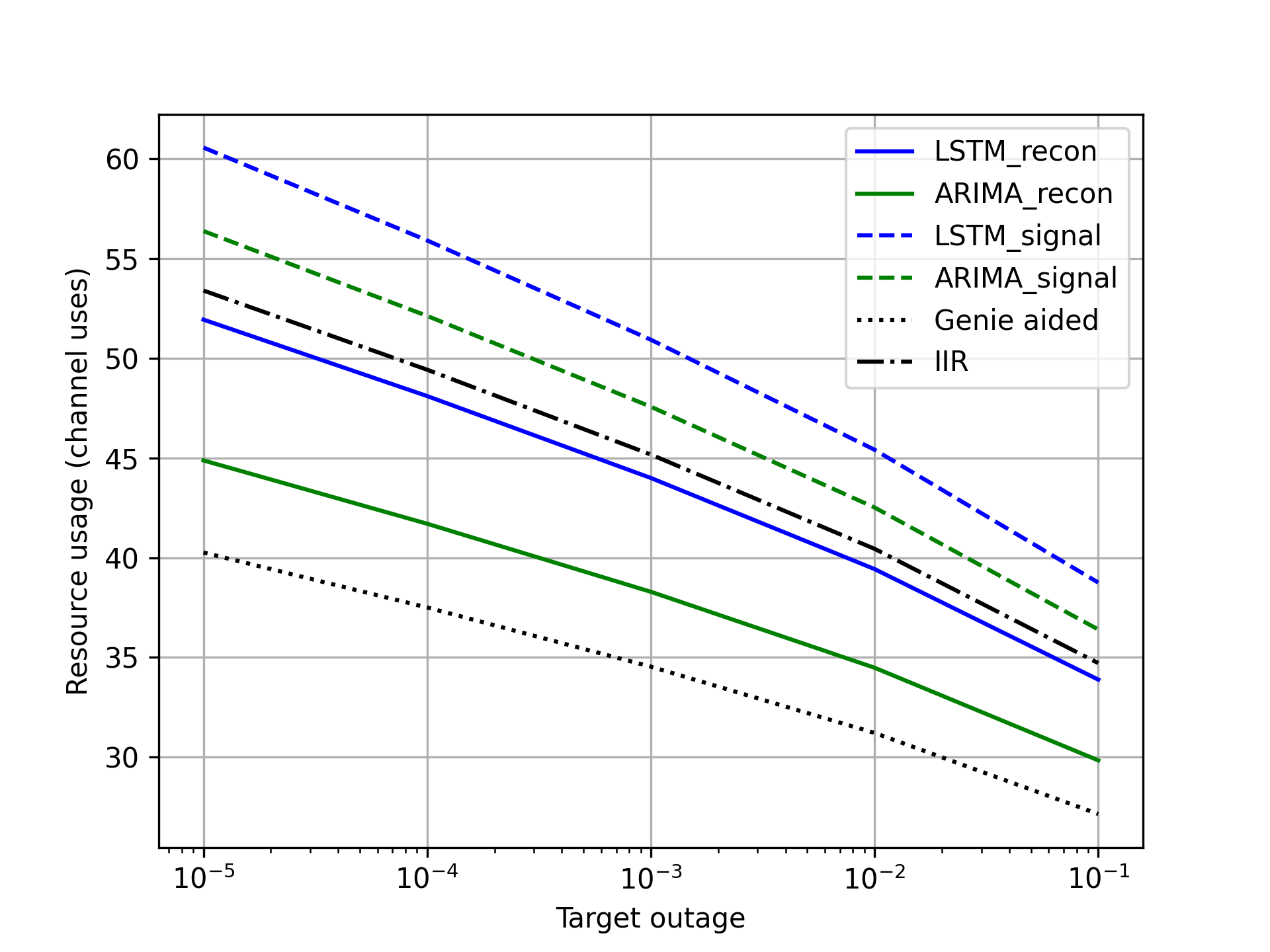}
    \caption{Resource usage versus target outage for different prediction methods.}
    \label{fig:resource_usage}
\end{figure}

We can observe that resource allocation without EMD utilizing only LSTM, and ARIMA methods shows poor performance compared to genie-aided resource allocation but perform better than IIR-based resource allocation. In contrast, achieved outages using EMD-based prediction methods demonstrate better performance than IIR filter-based methods and methods that do not use EMD. The resource allocation method, which uses EMD and ARIMA for interference prediction, has allocated resources in a nearly optimal way to meet almost target error rates, as shown in the $ARIMA recon$ curve. Then, the second-best resource allocation has been achieved with the EMD-based LSTM method, as shown in the $LSTM recon$ curve. It also proves the EMD-based prediction methods could allocate resources more efficiently than the non-EMD prediction methods. The resource usage corresponding to the above outage targets is shown in Fig.~\ref{fig:resource_usage}. The genie-aided estimator has allocated an optimal minimal number of resources because it is assumed to have exact prior knowledge about the interference. Our proposed EMD-based prediction schemes allocate resources to near-optimum performance, which assigns fewer resources than state-of-the-art interference prediction schemes.

\section{Conclusion}
\label{sec:conclu}
In this paper, predictive resource allocation for URLLC services was investigated. The desired downlink in the presence of $N$ interfering links was considered. We used the EMD algorithm to decompose the total interference signal power into IMFs and residuals. This decomposition allowed us to predict future interference values precisely. Along with the EMD, two prediction algorithms were used to train the models: LSTM and ARIMA. While heading toward the final objective of efficient resource allocation, the performance of EMD based hybrid prediction scheme was also evaluated. Based on the predicted interference values, downlink resources were allocated to withstand the presence of actual interference. Finally, the performance of the proposed scheme was evaluated against the two baseline schemes and was able to achieve near-optimal performance.

\bibliographystyle{IEEEtran}
\bibliography{main.bib}

\end{document}